%% file: TacisSNS.tex
\documentclass[manuscript]{acmart}

\renewcommand\footnotetextcopyrightpermission[1]{}

\usepackage{booktabs}
\usepackage{tabularx}
\usepackage{cleveref}
\usepackage{float}
\usepackage{enumitem}
\usepackage{soul}
\usepackage{placeins}
\usepackage{pifont}

\AtBeginDocument{%
  }

\acmConference[CHI '26]{CHI Conference on Human Factors in Computing Systems}{April 13--17, 2026}{Barcelona, Spain}

\setcopyright{none}

\begin{document}

\title[Trace-Aware Workflows for Co-Creating Branded Content with Generative AI]{Trace-Aware Workflows for Co-Creating Branded Content with Generative AI} %\fy{I feel it's not AI-assisted but human-assisted AI generation}}

\author{Taehyun Yang}
\affiliation{%
  \institution{University of Maryland}
  \city{College Park}
  \country{USA}}
\email{taeyang@umd.edu}

\author{Eunhye Kim}
\affiliation{%
  \institution{KAIST}
  \city{Daejeon}
  \country{Republic of Korea}
}
\email{gracekim027@kaist.ac.kr}

\author{Zhongzheng Xu}
\affiliation{%
  \institution{University of Maryland}
  \city{College Park}
  \country{USA}
  }
\email{zxu169@umd.edu}

\author{Fumeng Yang}
\affiliation{%
  \institution{University of Maryland}
  \city{College Park}
  \country{USA}
  }
\email{fy@umd.edu}

\renewcommand{\shortauthors}{}

\begin{abstract}

Generative AI tools have lowered barriers to producing branded social media images and captions, yet small-business owners (SBOs) still struggle to create on-brand posts without access to professional designers or marketing consultants. Although these tools enable fast image generation from text prompts, aligning outputs with a brand’s intended look and feel remains a demanding, iterative creative task. In this position paper, we explore how SBOs navigate iterative content creation %\fy{I really don't think creative decision is the right word in this paper: it means the decision is creative}
and how AI-assisted systems can support SBOs' content creation workflow. We conducted a preliminary study with 12 SBOs who independently manage their businesses and social media presence, using a questionnaire to collect their branding practices, content workflows, and use of generative AI alongside conventional design tools. We identified three recurring challenges: (1) translating brand ``feel'' into effective prompts, (2) difficulty revisiting and comparing prior image generations, and (3) difficulty making sense of changes between iterations to steer refinement. Based on these findings, we present a prototype that scaffolds brand articulation, supports feedback-informed exploration%\fy{gpt really like something like X-driven}
, and maintains a traceboard of branching image iterations. Our work illustrates how traces of the iterative process can serve as workflow support that helps SBOs keep track of explorations, make sense of changes, and refine content.

\end{abstract}




\begin{teaserfigure}
  \centering
  \includegraphics[width=\textwidth]{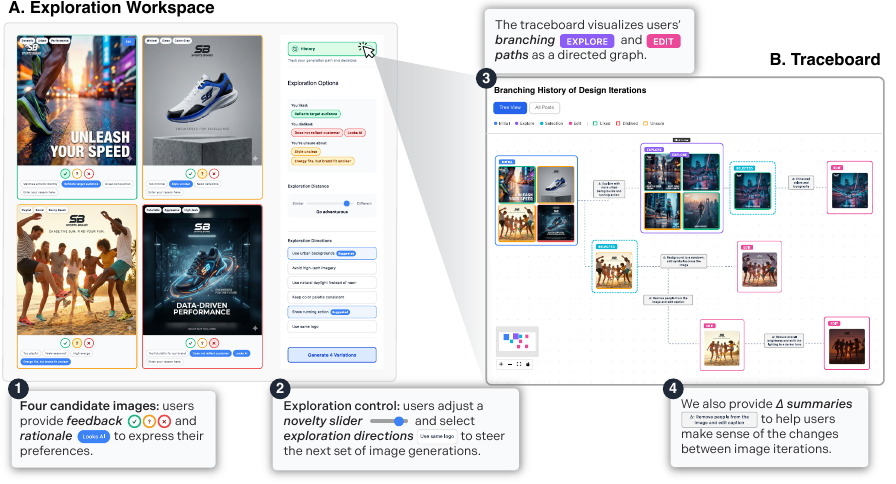}
  \vspace*{-20pt}
  \caption{Overview of our prototype for branded content creation: \textbf{(A) a creative exploration workspace}, where users generate, evaluate, and refine branded images, and \textbf{(B) a traceboard}, which traces and visualizes the history of image generations and edits. In the exploration workspace, {\color{annotationcolor}\ding{182}} users provide lightweight feedback (like/dislike/unsure along with short rationales) to express preferences towards individual images, and  {\color{annotationcolor}\ding{183}} use AI-guided exploration controls (suggested directions and a novelty slider) to steer the next set of image generations. The traceboard  {\color{annotationcolor}\ding{184}}  visualizes users' branching exploration and edit paths as a directed graph while  {\color{annotationcolor}\ding{185}}~change summaries ($\Delta$) form the edges of the graph describing what was modified between iterations, helping users make sense of the changes between iterations.}
  \Description{Overview of a trace-aware interface for AI-assisted branded content creation. The left panel shows a creative exploration workspace where a user reviews multiple generated promotional images, provides like/dislike/unsure feedback and short rationales, and uses suggested edit directions and a novelty slider to generate new variations. The right panel shows a traceboard: a branching node-link graph of image iterations and edits, with highlighted branches and natural-language change summaries ($\Delta$) attached to nodes to explain differences between successive versions.}
  \label{fig:teaser}
\end{teaserfigure}

\input{tricks}

\maketitle

\clubpenalty=-1000      % allow orphans (first line alone at bottom)
\widowpenalty=-1000     % allow widows (last line alone at top)
\displaywidowpenalty=-1000
\raggedbottom

\input{sections/1_introduction}
\input{sections/2_preliminarystudy}

\input{sections/3_features}
\input{sections/4_discussion}

%\fy{it's a good practice to include DOI in references whenever you can. It would also be easier for reviewers to verify the references}

\bibliographystyle{ACM-Reference-Format}
\bibliography{references}

\appendix
\input{sections/A_appendix}

\end{document}

%% file: tricks.tex
% 
% ------------------------------------------------------------------------------------------------ 
% this is where we disable/enable the notes
\newif\ifnotes
\notestrue % display notes
%\notesfalse % hide notes
% ------------------------------------------------------------------------------------------------  

\definecolor{annotationcolor}{HTML}{000000}

\definecolor{fycolor}{HTML}{8654d1}
\newcommand{\fy}[1]{\ifnotes{\leavevmode\color{fycolor}{(Fumeng: #1)}}\fi}
\newcommand{\fyreplace}[2]{\ifnotes{\color{gray}{\st{#1} \color{fycolor}{(Fumeng: #2)}}}\else{#2}\fi}
\newcommand{\fyedit}[1]{\ifnotes{\color{fycolor}{#1}}\else{#1}\fi}
\newcommand{\fydelete}[1]{\ifnotes{\color{gray}{\st{#1}}}\fi}
\newcommand{\fyadd}[1]{\ifnotes{\color{fycolor}{#1}}\else{#1}\fi}

%% file: sections/1_introduction.tex
\section{Introduction}

Small businesses comprise roughly 90\% of businesses worldwide \cite{worldbank2025sme,un2025msme}. 
 For a local bakery owner, platforms like Instagram, Facebook, and LinkedIn have become the first place a potential customer encounters their brand, forms an impression, and decides whether to engage. 
The content shared on these social media platforms, such as promotional images, branded graphics, and visual assets,   can directly influence brand reputation and influence purchasing \cite{kraus2019content, taiminen2015usage}. Therefore, a consistent, professional presentation is a competitive necessity for these social media contents.

Yet most small business owners (SBOs) struggle to meet this expectation. Unlike large enterprises with dedicated marketing teams, SBOs typically operate without the time, budget, or design expertise needed to produce professional visual content \cite{alford2018marketing}. Tasks that experienced designers and brand strategists handle systematically fall entirely on the business owner themselves, including defining a coherent visual direction, maintaining aesthetic consistency across posts, and ensuring every asset feels ``on brand.''

%\fy{I separate them, too much going on in one par}

% SBOs are increasingly responsible for producing captioned promotional images, branded graphics and visual assets for platforms like Instagram, Facebook, and LinkedIn \cite{ashley2015creative, kozinets2008wisdom, arvidsson2016brand}. 

% Recent Generative AI tools such as DALL-E \cite{openai_dalle}, Midjourney \cite{midjourney}, and Canva \cite{canva_ai} enable non-experts to produce visual content through text prompts \cite{zamfirescu2023johnny,li2024user,weisz2024design}.
% These tools have shown promising results in generating complex, real visuals. 
% Many SBOs have adopted these tools, hoping to reduce production costs and time \cite{karnatak2025acai,liu2025influencer,verizon2025smb}. However, creating visual is often an interactive, complex process due to their imperfection and the difficult of articulating needs. SBOs must typically refine visual direction, explore alternatives between multiple images, and align outputs with brand identity over multiple attempts.
% These Generative  AI tools, per se, lack support for the iterative process that is required to create effective SM content, creating an oppontunities for specified interface to 
% Given SBOs' time constraints and the iterative nature of brand-aligned creation,

Recent generative AI tools such as Midjourney \cite{midjourney} and DALL-E \cite{openai_dalle}, present both opportunities and challenges for SBOs. They have enabled non-experts to produce visual content through natural language prompts~\cite{zamfirescu2023johnny,li2024user,weisz2024design}, showing promising results in generating complex, high-quality visuals, and many SBOs have adopted them hoping to reduce production costs and time~\cite{karnatak2025acai,liu2025influencer,verizon2025smb}. However, creating effective visual content is rarely a one-shot process. Due to the inherent imprecision of generative AI outputs and the difficulty of articulating intents, SBOs typically iterate extensively. However, current generative AI tools offer little support for this iterative process, providing no structure for organizing explorations, externalizing brand preferences, or converging on a coherent visual identity over time. This very gap presents an opportunity for specialized interfaces that scaffold the iterative creation process and help SBOs articulate and refine their brand identity.

In this paper,  we propose a system that preserves exploration history and makes iterations easy to compare. By making prior attempts and their outcomes visible, the design helps SBOs avoid losing promising directions and assess whether each step moves closer to their intended brand direction. %\fy{gpt really likes X-aware as well. The previous version tries to do too much while this is clearly a topic sentence for our work}
We first conducted a questionnaire with 12 SBOs to understand the challenges they face during their process of creating social media content. 
%\fydelete{From these findings, we derive a set of design considerations for supporting traceable, iterative content creation.}\fy{we deleted this} 
Guided by the insights discovered, we introduce a prototype system that scaffolds brand articulation, preserves branching exploration histories, and makes generative changes interpretable through structured creativity traces. Through this work, we contribute a trace-aware approach %\fy{did we have a design space?}
to support users working with generative systems and highlight how interaction traces can be served as creative scaffolds for understanding iterative design practices.\looseness=-10 %\fy{avoid orphans}

%% file: sections/2_preliminarystudy.tex
\section{Preliminary Study}
\label{sec:prelim}
%We conducted a preliminary study to understand the challenges SBOs encounter while creating SM content using generative AI systems alongside conventional design tools (e.g., Illustrator, Figma, Canva, etc.). \fy{move to intro}
%Our goal was to examine how SBOs articulate brand intent, navigate creative decisions, and manage iterative content workflows without formal design training. \fy{describe how this goal is connected to your overall goal -- support SBOs}
%We recruited twelve SBOs who independently operated their businesses and conducted a semi-structured interview about their content workflows, definition of brand identity, and tool use. \fy{I thought it's a questionnaire?} \fy{refer to your table, like their demographics can be found in Table X} \fy{add this study has been approved by our institutional IRB. }

%The study used a semi-structured interview about SBOs’ SM content workflows, brand identity, and tool use. We asked about (1) challenges, goals, and high-stress moments in SM management, (2) how they define and recognize “on-brand” versus misaligned content, including difficulties expressing brand “vibe” in words or visuals, and (3) experiences with GenAI tools for content creation, including perceived benefits, concerns about brand misrepresentation, and desired features for future tools.

% To better support SBOs in AI-assisted content creation through traceable creative processes, 
%\fy{`conducted a questionnaire' is weird}
We first conducted a formative study to understand the specific challenges SBOs face when creating social media content. We recruited 12 SBOs who independently operated their businesses and administered a questionnaire on their content workflows, definition of brand identity, and tool use with generative AI systems alongside conventional design tools (e.g., Illustrator, Figma, Canva, etc.).  This study was approved by our  IRB with participant demographics available in Table 1 (Appendix~\ref{app:participants}). After thematic analysis, we identify three challenges, presented as follows. 
%\input{Tables_Figures/table}

%\input{Tables_Figures/smalltable}

% \subsection{Findings} \fy{not nessary to have a single subsection}

% Participants reported three recurring limitations when creating SM content.

%Second, participants \textbf{handled end-to-end content production as an iterative but poorly structured process}. Without the support of professional teams, SBOs were responsible for every stage--from ideation and design to captioning and publishing--
%\fy{could delete the previous sentence if it's mentioned in intro}
%and repeatedly cycled through generating, revising, and evaluating content. 
%However, this iterative work left little persistent structure: 
%Their ideas evolved across tools, intermediate artifacts were easily lost, and prior directions were difficult to revisit or compare. 
%As a result, participants struggled to maintain a clear sense of how their content developed over time. \fy{the three sentences above could be compressed into 2}
%P3 explained: \textit{``I move between three apps to see which version is better. It’s a hassle to try different captions and images across tools, and it’s hard to keep track.''}

 \textbf{Participants struggled to translate implicit brand knowledge into effective prompts (C1)}. Business owners often had a well-formed sense of how their brand should feel but lacked the vocabulary or design expertise to express it in actionable terms (e.g., visuals, tone, composition). This knowledge was experience-based, and articulating it required additional effort within already time-constrained workflows. P12 shared: \textit{``I didn’t know how to put that warmth into words, so I ended up not posting it.''} Because this translation gap prevents SBOs from effectively prompting AI tools, systems should provide scaffolds and examples that help convert experience-based intuitions into prompts.

\textbf{Participants handled end-to-end content production as an iterative but poorly structured process (C2)}. SBOs repeatedly cycled through generating, revising, and evaluating content. Very often, their ideas evolved across tools but intermediate artifacts were easily lost, making prior directions difficult to revisit or compare. P3 explained: \textit{``I move between three apps to see which version is better. It's a hassle to try different captions and images across tools, and it's hard to keep track.''} Because this lack of structure makes it difficult to evaluate progress or return to other design alternatives, systems should maintain persistent records of exploration.

%Third, \textbf{AI generation often outpaced SBOs' ability to interpret what changed across iterations}. Because they lacked formal design training and worked without collaborators to validate decisions, participants found it difficult to understand why outputs shifted or whether those shifts aligned with their brand. 
%\fy{whether those shifts aligned with their brand. - is this true? This seems contradictory to "Business owners often had a well-formed sense"}
%Small edits \fy{small edits in prompt?} could produce large changes in style or tone without clear explanations, making progress difficult to evaluate or steer. P4 noted: \textit{``...the AI takes a giant leap when I only asked for a small step. I'll change one word and the whole vibe flips. I can't tell if I'm getting closer.''}\looseness=-10

Finally, \textbf{SBOs struggled to judge whether iterations were moving closer to their intended style and tone~(C3)}. Because they lacked formal design training and worked without experts, participants found it difficult to evaluate whether each new iteration moved closer to their intended style and tone. Small prompt edits could produce large changes in style or tone without clear explanations, making progress difficult to evaluate or steer. P4 noted: \textit{``...the AI takes a giant leap when I only asked for a small step. I'll change one word and the whole vibe flips. I can't tell if I'm getting closer.''} As such, systems should make changes between iterations more explicit to support the content exploration and selection. \looseness=-10%\fy{avoid orphans}% \fy{creative decision-making means the decision-making process is creative} \fy{I'm not sure whether interpretable is the correct word. After reading the whole thing, I may use "sense-making" "make sense of "}

%\fy{For such a short paper, Design objectives can be merged with prototype features, especially your paragraph headers below are just repeating these DGs. Also, I think they are too specific to be guidelines. }

% \subsection{Design Guidelines}

% Based on the findings, we derive three design guidelines to better support AI-mediated creative exploration for SBOs.

% \begin{itemize}[leftmargin=10pt,label={},topsep=0pt]
%     \item \textbf{DG1. Scaffold the Translation of Brand Knowledge into Actionable Inputs.}
% Provide structured prompts, examples, and feedback that help SBOs convert experience-based brand judgments into concrete instructions without requiring professional design knowledge.
%     \item \textbf{DG2. Preserve Process Visibility Across Iterations.}  
% Provide a centralized workspace that maintains a persistent record of evolving ideas, allowing SBOs to branch, revisit, and refine content without losing intermediate artifacts.\looseness=-10
%     \item \textbf{DG3. Support Solo Creative Exploration Through AI-Recommended Directions.}  
% Offer proactive edit suggestions and exploratory prompts that help SBOs discover creative directions independently. Allow simple feedback on generated variations so the system can iteratively refine toward brand-fit, replacing the collaborative ideation and validation typically provided by design or marketing teams.

% \end{itemize}

%% file: sections/3_features.tex
\section{Prototype Features}
%\begin{figure}[t]
%    \centering
%    \includegraphics[width=\linewidth]{Tables_Figures/figure2.png}
%    \caption{Workflow of our prototype.}
%    \Description{A workflow diagram illustrating the system pipeline for supporting brand externalization, including structured prompting, trace capture, and reflection.}
%    \label{fig:workflow}
%\end{figure}

 %\fydelete{Based on the findings, we discuss three features to better support AI-mediated creative exploration for SBOs.}\fy{AI-assisted in the title} 
 
 Based on the findings, we discuss four features to better support SBOs in creating social media content with generative AI tools.
 We implemented these features in a prototype system (see \Cref{fig:teaser}) using {Gemini 3 Pro Image Preview} (``Nano Banana'') for image generation, \texttt{GPT-4o-mini} for prompt auto-suggestions and edit recommendations, and \texttt{GPT-4o} for generating natural-language summaries of changes between iterations.% \fy{maybe briefly mention the models used here}

% \subsection{Supporting Brand Externalization}
\textbf{Feature 1. Scaffolding Brand Knowledge Through Structured Prompting.}
Users begin by entering core brand information, including the brand name, category, and a textual brand description. 
%\fydelete{Our preliminary study found that SBOs often have well-formed brand intuitions but struggle to articulate them in ways that guide content creation.}\fy{this is on the same page}
To help SBOs express their brand intuitions (C1) in prompts, the system provides AI-assisted autocomplete that helps SBOs externalize knowledge without requiring professional marketing knowledge (\Cref{fig:appendix_autocomplete}). %\fy{Appendix has a different numbering system to avoid confusion. }
Autocomplete operates in two modes. 
First, at sentence onset, the system detects missing brand dimensions using a schema inspired by Kapferer's Brand Identity Prism ~\cite{kapferer2008new} and offers scaffolding topics that frame brand knowledge in accessible terms (e.g., when information about uniqueness is absent, the system suggests ``Something that makes [brand name] different from competitors is...''). 
Second, during mid-sentence pauses, the system offers context-sensitive phrase completions that help SBOs articulate implicit feelings (e.g., after ``When customers interact with my brand, I want them to feel...,'' the system proposes emotional descriptors based on prior context). A hoverable indicator visualizes which portions of the description correspond to specific identity facets, helping users identify gaps in articulation and surface overlooked aspects of their brand understanding.%\fy{Ideally, I'd suggest embedding all buttons, colors, etc., and stylish challenges and features. But I let it go...}

\textbf{Feature 2. Translating Brand Judgments into Visual Directions.}
Articulating brand knowledge, however, does not guarantee visually aligned outputs. To help SBOs translate brand descriptions into visuals that match their intended feel (C1), the system supports iterative evaluation and refinement of AI-generated images. After the brand description is completed, the system generates four candidate images conditioned on the provided information. Users evaluate each image through their existing brand knowledge, providing lightweight feedback (like, dislike, unsure; see ~\Cref{fig:teaser}A) accompanied by automatically inferred rationales from GPT-4o (e.g., ``Too futuristic for our brand''). Users may revise these rationales using their own language. Selecting \textit{Generate 4 Variations} generates new images that incorporate this feedback, with a novelty control allowing users to balance between diverse exploration and incremental refinement (\Cref{fig:teaser}A). When users identify a promising direction, they can further refine images through prompt-based regeneration with suggested edit instructions or region-specific modifications.%\fy{also link these back to C1-C3?}

\textbf{Feature 3. Preserving Process Visibility Through Creativity Traces.}
To address the lack of structure in SBOs' iterative workflows (C2), the system provides a centralized traceboard (\Cref{fig:teaser}B) that maintains a record of the entire content development process.
%\fydelete{Our preliminary study found that SBOs cycle through generating, revising, and evaluating content across multiple stages but this iterative work was frequently unstructured. }
The traceboard records all generation and refinement actions (e.g., \textit{Explore}, \textit{Edit}, \textit{Selection}) as creativity traces, visualized as a directed graph in which nodes represent artifacts (images) and edges represent transformations (prompt changes, feedback updates, edits). This structure preserves branching exploration trajectories from the initial set, allowing users to see how their content evolved over time rather than losing track across iterations.
Users can select any node to revisit prior states, compare alternatives side-by-side, or resume exploration from earlier points without losing progress. 
When users identify a promising direction but later want to explore a different approach, they may revise feedback at any earlier node to regenerate variants from that branch while maintaining the full context of what they've already tried. 
%\fydelete{This addresses the difficulty of moving between versions across tools by keeping all exploration visible and accessible within a single workspace.}

\textbf{Feature 4. Supporting Solo Creative Exploration Through AI Guidance.}
To help SBOs working independently understand AI-generated changes and steer refinement (C3), the system offers creative suggestions and makes generative changes visible.
First, the system proactively suggests exploratory directions in the \textit{Edit} function, recommending targeted edit prompts (e.g., ``try a warmer color palette'') that guide refinement without requiring design vocabulary (\Cref{fig:appendix_edit}). 
These suggestions help users quickly specify refinement directions without needing design expertise or extensive prompt iteration.
Second, the system makes each iteration understandable by revealing what changed between generations. For each generated image, the system extracts descriptive keywords that summarize visual attributes in natural language (\Cref{fig:teaser}A). When users provide  feedback, the system guides subsequent \textit{Explore} generations toward preferred characteristics while avoiding undesired ones. The system then summarizes differences between successive outputs and attaches these natural language change descriptions to the corresponding traceboard nodes (\Cref{fig:teaser}B), helping users evaluate brand fit.

%% file: sections/4_discussion.tex
\section{Workshop Alignment and Future Work}
Our work aligns with the workshop's goal of advancing trace analysis by proposing how interaction traces for AI-assisted content creation can be captured beyond simple prompt histories. Tracing interactions that were previously unrecorded and transient, we represent SBOs' iterative processes as structured traces---recording feedback (and rationales), edits, branching alternatives, and natural language summaries of change---to reflect how SBOs refine visual direction and assess brand fit over time. Our traceboard organizes these traces as a directed graph of generated artifacts and transformations, enabling inspection of exploration paths and iteration-to-iteration changes.

In doing so, we hope to provide a concrete trace representation that can serve as analyzable data for studying creative activity traces. The traceboard supports analysis at multiple levels, from local edits between single images to higher-level patterns across multiple image sets. In follow-up work, we plan to trace and analyze additional signals within these workflows (e.g., feedback rationales, edit types, and branching strategies) and examine how trace patterns (e.g., branching depth, revision frequency, and feedback granularity) relate to downstream outcomes such as perceived content quality and audience response to the resulting branded images.

%% file: sections/A_appendix.tex
\newpage

%\fy{references are part of the necessary paper and should be put before appendix}

\appendix
\renewcommand{\thefigure}{\Alph{section}.\arabic{figure}}
\setcounter{figure}{0}
\renewcommand{\thetable}{\Alph{section}.\arabic{table}}
\setcounter{table}{0}

%\fy{Also, clearly mark the beginning of appendix; use different numbering systems for appendix}

{\noindent\bf\large{\textsf{Appendices}}}

\section{Participant Demographics}
\label{app:participants}
\input{Tables_Figures/table} % or paste the table here

% \fy{make sure tables are in the corresponding when you don't plan to refer to them}

\FloatBarrier

\section{Additional System Details}
\label{app:system}

\subsection{Prompt Autocomplete}
The Prompt Autocomplete system scaffolds brand articulation during content setup. It suggests structured prompts based on missing brand dimensions and provides context-aware phrase completions to help users express brand attributes in actionable terms.

\begin{figure}[H]
  \centering
  \includegraphics[width=0.43\linewidth]{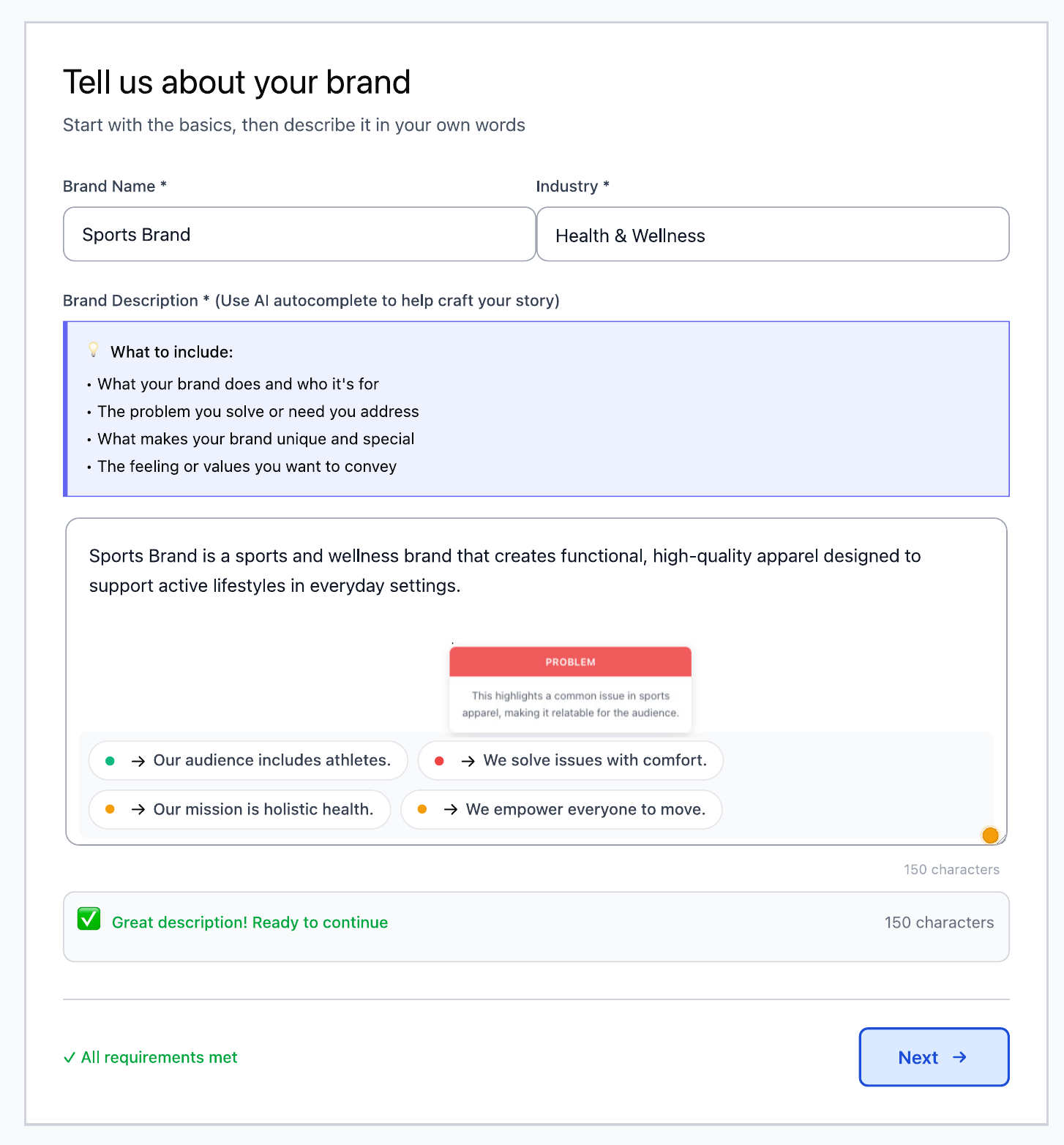}
  \caption{Prompt autocomplete system. The interface suggests structured brand dimensions and context-aware phrase completions to help users articulate brand intent.}
  \Description{Screenshot of the prompt autocomplete interface showing a brand description text box with suggested sentence starters and phrase completions linked to brand identity facets.}
  \label{fig:appendix_autocomplete}
\end{figure}

\subsection{Edit Interface}
The Edit interface supports targeted refinement of generated images after an initial exploration round. Users can apply suggested edit prompts (e.g., color adjustments, stylistic shifts), modify text instructions directly, or regenerate specific regions of an image.

\begin{figure}[H]
  \centering
  \includegraphics[width=\linewidth]{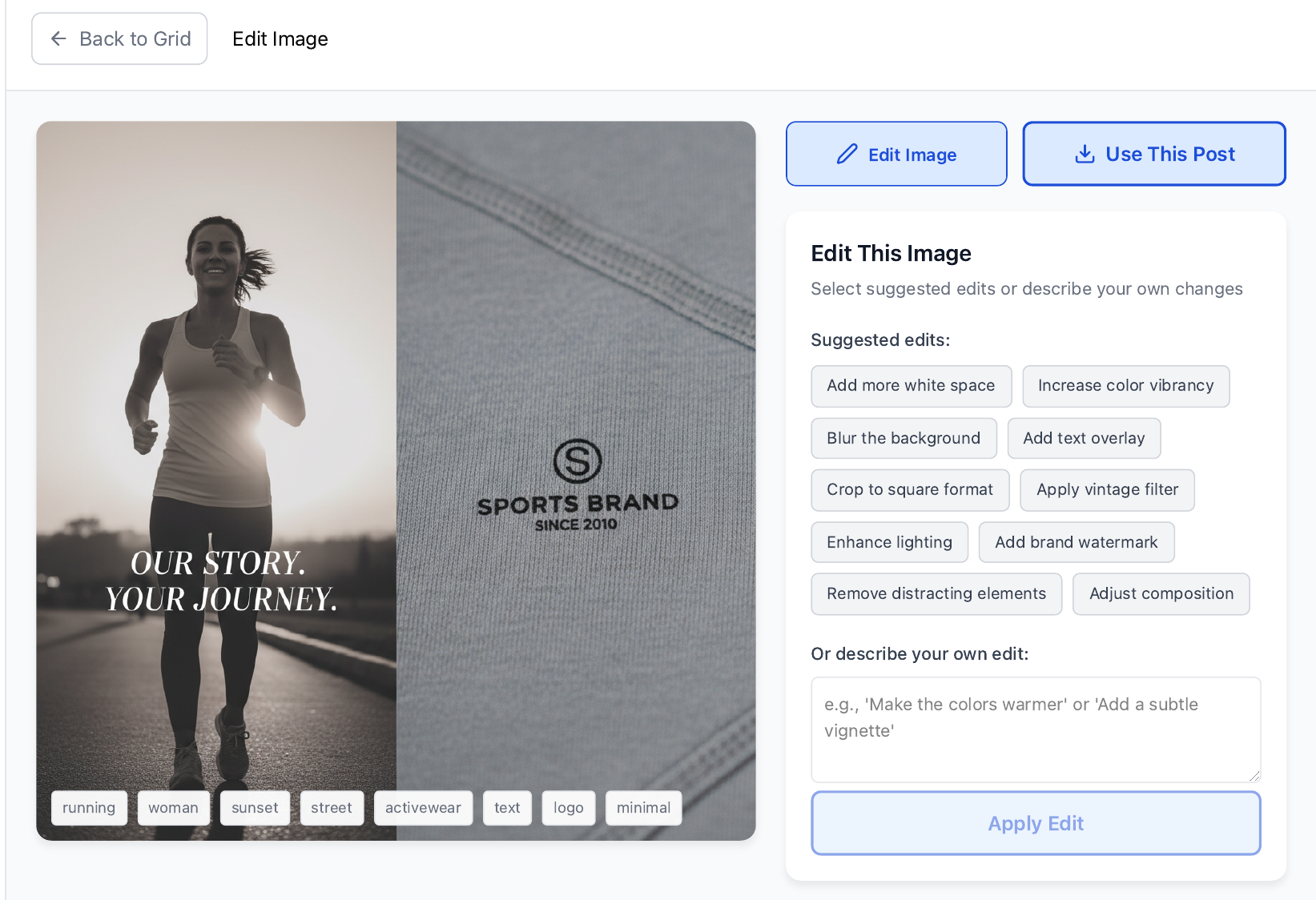}
  \caption{Edit interface for targeted refinement. Users can apply AI-suggested edit directions or modify prompts directly; each edit creates a new branch in the traceboard.}
  \Description{Screenshot of the edit interface showing a selected generated image, suggested edit directions, editable prompt text, and controls to apply refinements that create new branches in the traceboard.}
  \label{fig:appendix_edit}
\end{figure}

%% file: Tables_Figures/table.tex
\begin{table}[h]
\caption{Participant demographics in our preliminary study. Business age indicates how long participants have operated their brand. Abbreviations: IG=Instagram, YT=YouTube, TT=TikTok, LI=LinkedIn.}
\label{tab:participants}
\centering
{\small
\setlength{\tabcolsep}{4pt}
\begin{tabularx}{\linewidth}{l l l X l}
\toprule
ID & Role & Business age & Industry/Product & Platforms \\
\midrule
P1  & Aspiring entrepreneur & 6 mo  & Fashion & IG, YT \\
P2  & Small-business owner  & 2 yr  & Food Business & IG, Blog \\
P3  & Side-business owner   & 3 yr  & Publishing / branding & IG, TT \\
P4  & Side-business owner   & 4+ yr & Design consulting & IG, LI \\
P5  & Small-business owner  & 3 yr  & Food Business & IG, Kakao \\
P6  & Side-business owner   & 2 yr  & Fashion (sportswear) & IG, YT \\
P7  & Small-business owner  & 3 yr  & Education & IG, Blog \\
P8  & Small-business owner  & 1 yr  & Fashion & IG, Blog, Kakao \\
P9  & Side-business owner   & 1 yr  & Beauty & IG, Kakao, Band \\
P10 & Small-business owner  & 2 yr  & Cafe & IG, Blog \\
P11 & Freelancer            & 1 yr  & Music production & IG \\
P12 & Small-business owner  & 6 mo  & Cafe & IG, TT \\
\bottomrule
\end{tabularx}
}
\end{table}

%% file: references.bib
@misc{verizon2025smb,
  author       = {{Verizon Business}},
  title        = {2025 State of Small Business Survey: Surge in AI, Cybersecurity and Social Media Demand},
  year         = {2025},
  month        = {May},
  howpublished = {\url{https://www.verizon.com/about/news/2025-state-of-small-business-survey}},
  note         = {Accessed: 2026-02-06}
}

@article{kraus2019content,
  title={Content is king: How SMEs create content for social media marketing under limited resources},
  author={Kraus, Sascha and Gast, Johanna and Schleich, Moritz and Jones, Paul and Ritter, Michael},
  journal={Journal of Macromarketing},
  volume={39},
  number={4},
  pages={415--430},
  year={2019},
  publisher={SAGE Publications Sage CA: Los Angeles, CA},
  doi = {10.1177/0276146719882746}
}

@article{taiminen2015usage,
  title={The usage of digital marketing channels in SMEs},
  author={Taiminen, Heini Maarit and Karjaluoto, Heikki},
  journal={Journal of Small Business and Enterprise Development},
  volume={22},
  number={4},
  pages={633--651},
  year={2015},
  publisher={Emerald Group Publishing Limited},
  doi = {10.1108/JSBED-05-2013-0073}
}

@incollection{alford2018marketing,
  title={Marketing technology for adoption by small business},
  author={Alford, Philip and Page, Stephen John},
  booktitle={Social Media and Interactive Communications},
  pages={99--113},
  year={2018},
  publisher={Routledge},
  doi= {10.1080/02642069.2015.1062884}
}

@inproceedings{liu2025influencer,
  title     = {Influencer: Empowering Everyday Users in Creating Promotional Posts via AI-infused Exploration and Customization},
  author    = {Liu, Xuye and Sun, Annie and An, Pengcheng and Ma, Tengfei and Zhao, Jian},
  booktitle = {Proceedings of the 2025 CHI Conference on Human Factors in Computing Systems},
  year      = {2025},
  publisher = {Association for Computing Machinery},
  address   = {New York, NY, USA},
  pages     = {1--19},
  doi = {10.1145/3706598.3713309}
}

@misc{karnatak2025acai,
  title         = {ACAI for SBOs: AI Co-creation for Advertising and Inspiration for Small Business Owners},
  author        = {Karnatak, Nimisha and Baranes, Adrien and Marchant, Rob and Butler, Triona and Olson, Kristen},
  year          = {2025},
  howpublished  = {arXiv},
  eprint        = {2503.06729},
  archivePrefix = {arXiv},
  doi = {10.48550/arXiv.2503.06729}
}

@book{kapferer2008new,
  title     = {The New Strategic Brand Management: Creating and Sustaining Brand Equity Long Term},
  author    = {Kapferer, Jean-No{\"e}l},
  year      = {2008},
  publisher = {Kogan Page},
  address   = {London, UK}
}

@inproceedings{zamfirescu2023johnny,
  title={Why Johnny can’t prompt: how non-AI experts try (and fail) to design LLM prompts},
  author={Zamfirescu-Pereira, J Diego and Wong, Richmond Y and Hartmann, Bjoern and Yang, Qian},
  booktitle={Proceedings of the 2023 CHI conference on human factors in computing systems},
  pages={1--21},
  year={2023},
  doi={10.1145/3544548.3581388}
}

@inproceedings{weisz2024design,
  title={Design principles for generative AI applications},
  author={Weisz, Justin D and He, Jessica and Muller, Michael and Hoefer, Gabriela and Miles, Rachel and Geyer, Werner},
  booktitle={Proceedings of the 2024 CHI Conference on Human Factors in Computing Systems},
  pages={1--22},
  year={2024},
  doi= {10.1145/3613904.3642466}
}

@inproceedings{li2024user,
  title={User experience design professionals’ perceptions of generative artificial intelligence},
  author={Li, Jie and Cao, Hancheng and Lin, Laura and Hou, Youyang and Zhu, Ruihao and El Ali, Abdallah},
  booktitle={Proceedings of the 2024 CHI conference on human factors in computing systems},
  pages={1--18},
  year={2024},
  doi= {10.1145/3613904.3642114}
}

@misc{worldbank2025sme,
  author       = {{World Bank}},
  title        = {Small and Medium Enterprises (SMEs) Finance},
  year         = {2025},
  url          = {https://www.worldbank.org/en/topic/smefinance},
  note         = {Accessed: 2026-02-07}
}

@misc{un2025msme,
  author       = {{United Nations}},
  title        = {Micro-, Small and Medium-sized Enterprises Day},
  year         = {2025},
  url          = {https://www.un.org/en/observances/micro-small-medium-businesses-day},
  note         = {Accessed: 2026-02-07}
}

@misc{openai_dalle,
  author       = {{OpenAI}},
  title        = {{DALL·E}},
  howpublished = {\url{https://openai.com/dall-e}}
}

@misc{midjourney,
  author       = {{Midjourney}},
  title        = {{Midjourney}},
  howpublished = {\url{https://www.midjourney.com/}}
}
